# Error analysis of nuclear mass fits

J. Toivanen,[1,*] J. Dobaczewski,[1,2] M. Kortelainen,[1] and K. Mizuyama[1]

[1] *Department of Physics, P.O. Box 35 (YFL), FI-40014 University of Jyväskylä, Finland*
[2] *Institute of Theoretical Physics, University of Warsaw, ul. Hoża 69, PL-00-681 Warsaw, Poland.*

We discuss the least-square and linear-regression methods, which are relevant for a reliable determination of good nuclear-mass-model parameter sets and their errors. In this perspective, we define exact and inaccurate models and point out differences in using the standard error analyses for them. As an illustration, we use simple analytic models for nuclear binding energies and study the validity and errors of models' parameters, and uncertainties of its mass predictions. In particular, we show explicitly the influence of mass-number dependent weights on uncertainties of liquid-drop global parameters.

## I. INTRODUCTION

Mass or binding energy is one of the most fundamental properties of atomic nucleus. Measuring and modelling nuclear masses has been since many years, and still is, at the center stage of nuclear physics, see Ref. [1] for a recent review. Determination of mass from first principles, viz. quantum chromodynamics, is extremely difficult and only possible in lattice QCD for composite particles like mesons or nucleons [2], and is beyond anything possible or sensible for nuclei. For light nuclei, one can quite accurately calculate nuclear masses by using many-body technics that employ parametrized models of nucleon-nucleon (NN) and NNN interactions, see e.g. Ref. [3]. In these so-called *ab initio* models, parameters are partly fitted to other observables than mass (like NN phase shifts) and partly to masses (NNN interactions). There are many other, less sophisticated methods to calculate nuclear masses, and all of them include fitting to mass data to a larger or smaller extent. Therefore, there is an extensive history of mass fits in nuclear physics.

Nevertheless, and strangely enough, the history of error analyses of these mass fits is virtually nonexistent (but see notable examples in Refs. [4, 5]). As a consequence, there exist in the literature very many mass tables and mass predictions, but there are no estimates of the reliability of these results, which would be based on thorough methods of analyzing their uncertainties.

In the present study, we aim at (i) recalling the well-known methods that must be used to analyze errors along with any fits of parameters, and (ii) pointing several particular features of such analyses that are characteristic in applications to mass fits. At present, one cannot overestimate the importance of quantitatively analyzing the predictivity of mass calculations when applied to exotic nuclei far from stability. However, such mass calculations must be accompanied by predictions of their theoretical error bars. On the one hand, professional error analyses will put predictions on firm grounds—often showing explicitly that such predictions are simply impossible, when they are based on a given model fitted to a given set of masses. On the other hand, they will give quantitative information on how much a measurement of mass of the last available isotope (often very difficult) will improve predictivity of models.

As a benchmark number that characterizes mass fits, one has the mass root-mean-squared (rms) deviation, which nowadays does not go below about 0.6 MeV [1, 6, 7]. Down to this level, nuclear models were successfully used to describe nuclear masses, and moreover, they often correctly describe other observables like charge radii and other ground-state properties and excitations. In the present study we do not enter into the discussion of which observables, apart from mass, should be used to fit given models to data. Of course, error analyses should be performed when fitting any kinds of observables, although our particular example below concerns only a mass model.

In particular, the best Skyrme and Gogny energy-density-functional (EDF) methods [8], fitted to large numbers of nuclei, have resulted in rms deviations of 0.7–1.0 MeV from experimental masses. The deviations from experiment are not random, but show systematic patterns [9]. These patterns are a clear sign that the functionals are too simplified, see also Ref. [10]. Systematic methods are needed to improve EDF models by introducing new terms (for example, by using density-dependent coupling constants, see e.g. Refs. [11, 12], or higher-order derivative terms [13]) and testing the importance and physical feasibility of the new terms.

Current EDF models typically use 10–14 parameters or coupling constants. Skyrme functionals, for example have quite clear physical interpretation for all of the parameters of the functional. If the number of model parameters is drastically increased, the meaning and importance of parameters might not always be clear. To be able to understand the significance of each parameter, clear and efficient methods must be used, as discussed in the present study.

*jutato@phys.jyu.fi

## II. METHODS OF REGRESSION ANALYSIS

In this section we briefly recapitulate methods used in the standard linear regression method [14]. Along with presenting necessary definitions and main results, we also discuss several aspects that are specific to our particular problem of nuclear mass fits.

Let us assume that we use a model describing $j = 1, \ldots, m$ observables $e_j$ in terms $i = 1, \ldots, n$ parameters $x_i$, i.e.,

$$e_j = f_j(\vec{x}) \,. \tag{1}$$

To find an optimal set of parameters, a fitting procedure has to be used, whereupon the rms deviation (including in regression analysis a $1/(m-n)$ normalization)

$$\Delta^2_{\rm rms} = \frac{1}{m-n} \sum_{j=1}^{m} W_j \left( f_j(\vec{x}) - e^{\rm exp}_j \right)^2 \tag{2}$$

between experimental values of observables, $e^{\rm exp}_j$, and the observables given by model is minimized by adjusting the model parameters. This is called the least square fitting procedure. As is usually the case, the number of observables is larger than the number of parameters, $m > n$.

Each term in the sum of Eq. (2) is multiplied by a weight factor $W_j > 0$. In this respect we can single out two limiting situations, of an exact and an inaccurate model:

- The model of Eq. (1) is exact and deviations in Eq. (2) result solely from imprecisely measured experimental values. In this case, one takes the weights $W_j = (\Delta e_j)^{-2}$, where $\Delta e_j$ are experimental variances of observables $e_j$.

- The model of Eq. (1) is a poor approximation of reality and deviations in Eq. (2) are much larger than the experimental variances of observables. In this case, the choice of weights is quite arbitrary and can only be based on intuition. By using different weights one can, in fact, differentiate between importance of various observables in determining the model parameters. It is clear that the result of adjustment may crucially depend on the choice of weights.

In the nuclear mass fits discussed in the present paper, we are obviously in the case of an inaccurate model, by which typical experimental errors are of the order of a few tens of keV [15], but can also be as low as about 100 eV [16], while average deviations of mass models do not go below about 0.6 MeV [1]. In case of several different kinds of observables included in the fit, dependence of the results on weights is obvious, see e.g. recent comprehensive analysis in Ref. [5]. However, even if only nuclear masses are fitted, the 'natural' choice of weights, $W_j = 1$, is only a choice, and many other choices are possible, e.g. depending on whether one wants to put more weight into measured values of light or heavy, or stable or exotic nuclei. We illustrate this point in Sec. III below.

### A. Determination of parameters

The function (2) has an extremum when all its partial derivatives with respect to the model parameters $x_i$ are simultaneously zero,

$$\frac{\partial \left( \Delta^2_{\rm rms} \right)}{\partial x_i} = 0, \quad i = 1, \ldots, n \,. \tag{3}$$

These partial derivatives are in general non-linear functions of the model parameters; thus to get manageable equations, Eq. (1) has to be linearized, i.e.,

$$f_j(\vec{x}) \simeq f_j(\vec{x}_0) + \sum_{i=1}^{n} \left( \frac{\partial f_j}{\partial x_i} \right)_{\vec{x}=\vec{x}_0} (x_i - x_{0,i}) \,. \tag{4}$$

For observables related to total or single-particle energies, the non-linearities can actually be quite small [4, 10], but in general this is not a case and the linearized equations have to be solved iteratively.

We now introduce the notation that $\vec{x}_0$ is the set of parameters from previous iteration, by which $x_i - x^0_i$ is the change of parameters to be determined. We also denote the weighted deviations of observables from experiment by $y_j$,

$$y_j \equiv \sqrt{W_j} \left( e^{\rm exp}_j - f_j(\vec{x}_0) \right) \,, \tag{5}$$

and the weighted matrix of regression coefficients is denoted as

$$J_{ji} \equiv \sqrt{W_j} I_{ji} \tag{6}$$

for

$$I_{ji} = \left( \frac{\partial f_j}{\partial x_i} \right)_{\vec{x}=\vec{x}_0} \,. \tag{7}$$

Then, Eq. (2) can be written as

$$\Delta^2_{\rm rms} = \frac{1}{m-n} \sum_{j=1}^{m} \left( \sum_{i=1}^{n} J_{ji} (x_i - x^0_i) - y_j \right)^2 \,, \tag{8}$$

and Eq. (3) takes the form:

$$\left( J^T J \right) (\vec{x} - \vec{x}_0) = J^T \vec{y} \,. \tag{9}$$

It is now obvious that the parameters lying in the null space of $J^T J$ (if it is singular) cannot be determined. Moreover, during the fitting procedure it often happens that some parameters are very poorly determined by the experimental data. These parameters should be removed from the set because they have very large uncertainties and, if kept, would destroy the subsequent error analysis (see below). The poorly determined parameters can be found by first transforming to a new set of parameters, here called 'independent parameters' and then eliminating all non-important independent parameters from the fit.

This can be achieved by making a singular value decomposition (SVD) [17] of matrix $J$,

$$J_{ji} = \sum_{k=1}^{q} U_{jk} w_k V^T_{ki} \, , \tag{10}$$

where columns of the $m \times q$ matrix $U$ are orthogonal ($U^T U = 1$), columns of the $n \times q$ matrix $V$ are also orthogonal ($V^T V = 1$), and $q$ positive numbers $w_k$ are called singular values of $J$. Note that for singular matrix $J^T J$ one has $q < n$, and the vanishing singular values do not contribute to the sum in Eq. (10).

The SVD of $J$ allows one to calculate the inverse $\left(J^T J\right)^{-1}$ *outside* the null space of $J^T J = V w^2 V^T$,

$$\left(J^T J\right)^{-1} = V \frac{1}{w^2} V^T \, , \tag{11}$$

and the solution of Eq. (9) can now be expressed as

$$\vec{x} - \vec{x}_0 = \left(J^T J\right)^{-1} J^T \vec{y} = V \frac{1}{w} U^T \vec{y} . \tag{12}$$

The new independent parameters are now defined as $\vec{z} = V^T \vec{x}$. If some singular values become very small, the associated variables are simply dropped from Eq. (12), i.e.,

$$\begin{aligned} z_k - z_{0,k} &= \tfrac{1}{w_k} \textstyle\sum_{j=1}^{m} U^T_{kj} y_j \qquad \text{for} \;\; w_k > \epsilon \, , \\ &= 0 \qquad\qquad\qquad\qquad \text{for} \;\; w_k < \epsilon \, , \end{aligned} \tag{13}$$

and the new parameters $x_i$ become

$$x_i = x_{0,i} + \sum_{w_k > \epsilon} V_{ik} \frac{1}{w_k} \sum_{j=1}^{m} U^T_{kj} y_j \, . \tag{14}$$

These new values can now be used to continue iterations.

### B. Error estimates

After the iteration has converged, one can determine error estimates for the obtained parameters $x_i$. The method used here follows the standard multivariate regression analysis [18, 19] Assume that we take the scaled experimental observables and perturb them with a random noise that has zero mean value. The true experimental energies can now be thought of as being random variables but only one sample that has the values $\sqrt{W_j} e_j^{\rm exp}$ is known. The deviation of each model parameter $x_i$ from its mean can then be calculated from Eq. (12) as

$$x_i - \langle x_i \rangle = \sum_j \left( \left(J^T J\right)^{-1} J^T \right)_{ij} \left( y_j - \langle y_j \rangle \right) . \tag{15}$$

Then, the correlation matrix of parameters $x_i$ and $x_{i'}$ becomes

$$\langle \left(x_i - \langle x_i \rangle\right) \left(x_{i'} - \langle x_{i'} \rangle\right) \rangle = \sum_j \sum_{j'} \left( J \left(J^T J\right)^{-1} \right)_{ji} \left( \left(J^T J\right)^{-1} J^T \right)_{i'j'} \langle \left(y_j - \langle y_j \rangle\right) \left(y_{j'} - \langle y_{j'} \rangle\right) \rangle = \delta^2_{\rm rms} \left(J^T J\right)^{-1}_{ii'} \, , \tag{16}$$

where

$$\delta_{\rm rms} = t_{\alpha/2, m-n} \Delta_{\rm rms} \tag{17}$$

and $t_{\alpha/2,m-n}$ is Student's t-distribution [20] for $m-n$ degrees of freedom, necessary here because of small sample size. In Eq. (16) we have assumed that $y_j$ are *independent* random variables whose cross expectation values vanish and all have the same standard deviation, i.e.,

$$\langle \left(y_j - \langle y_j \rangle\right) \left(y_{j'} - \langle y_{j'} \rangle\right) \rangle = \delta_{jj'} \delta^2_{\rm rms} \, . \tag{18}$$

The average values of parameters, $\langle x_i \rangle$, are determined by the least square fitting procedure, $\langle x_i \rangle = x_{0,i}$. It is also assumed that the least square fitting gives an accurate estimate of the standard deviation of the observables $e_j$. With these assumptions, from Eq. (16) we get the following formula for the confidence interval of $x_i$ with $(1-\alpha)$ probability:

$$\Delta x_i \equiv \sqrt{\langle \left(x_i - \langle x_i \rangle\right)^2 \rangle} = \delta_{\rm rms} \sqrt{\left(J^T J\right)^{-1}_{ii}} \, . \tag{19}$$

It is now clear that small SVD values that appear in the inverse matrix of Eq. (11) spoil confidence intervals of all parameters, and have to be removed, as in Eq. (13). One should observe that Eq. (19) does implicitly depend on the weights through the definitions of Eqs. (5), (6), and (8).

We have to stress at this point that the error estimates of Eq. (19) have quite different meaning for the exact and inaccurate models discussed at the beginning of this section. In the first case, errors of parameters result solely from the statistical noise in measured

observables—variances thereof are supposed to be known and define weights in Eq. (2) as $W_j = (\Delta e_j)^{-2}$. Therefore, within the exact model, the assumption of equal variances, Eq. (18), is well justified. Such model then gives the minimum value of $\Delta^2_{\rm rms}$ near 1, which is the so-called $\chi^2$ test.

For an inaccurate model, the error estimates of Eq. (19) only give information on the sensitivity of the model parameters to values of the observables. They correspond to the situation where the experimental values are artificially varied far beyond their experimental uncertainties, so as to induce tangible variations in values of parameters. Eq. (18) then means that the range of this variation is inversely proportional to $\sqrt{W_j}$, i.e. it is commensurate with the importance attributed to a given observable. Here, the error estimates may depend on the weights, and thus are affected by their choices, similarly as the values of parameters are.

We are now in a position to discuss very important aspect of the mass fits, namely, the mass predictions and error propagation. Suppose that we apply the model of Eq. (1) not only to the measured masses but also to the masses of unknown nuclei,

$$\tilde{e}_j = f_j(\vec{x}) \,, \tag{20}$$

where the tilde means that the set of observables $\tilde{e}_j$ includes not only those used for the fit, $j = 1, \ldots, m$, but also many other ones, $j = m+1, \ldots, M$.

The error estimates of Eq. (19) allow us to estimate uncertainties of the predicted observables. With the same assumptions as before, but now using the parameters $x_i$ from the least square fit both for observables inside and outside the fitted set, we get

$$\left(\tilde{e}_j - \langle\tilde{e}_j\rangle\right)^2 = \sum_{ii'} \tilde{I}_{ji}\tilde{I}_{ji'} \left(x_i - \langle x_i\rangle\right)\left(x_{i'} - \langle x_{i'}\rangle\right) \,, \tag{21}$$

where $\tilde{I}_{ji}$ are the regression coefficients, Eq. (7), of observables $\tilde{e}_j$ with respect to the model parameters $x_i$. Then, the confidence intervals of predicted observables become

$$\Delta\tilde{e}_j = \sqrt{\langle\left(\tilde{e}_j - \langle\tilde{e}_j\rangle\right)^2\rangle} = \delta_{\rm rms}\sqrt{\left(\tilde{I}\left(J^TJ\right)^{-1}\tilde{I}^T\right)_{jj}} \,, \tag{22}$$

where we have used Eq. (16).

Equations (19) and (22) form the basis of the error analysis of our mass fits. The calculated error bars (19) of parameters $x_i$ must then be further scrutinized to analyze which parameters are necessary and which should be removed from the model. The confidence intervals (22) constitute estimates of predictivity of the model. Note that they should also be calculated for the observables that have actually been used in the fit. It is these intervals, and not the residuals $y_j/\sqrt{W_j}$, which have to be analyzed when discussing the quality of the model. Indeed, it is obvious that the residuals can be arbitrarily small for some observables, or for some types of observables (e.g., masses of semimagic spherical nuclei), while the model can still be quite uncertain in describing these same observables.

## III. EXAMPLE APPLICATION

To illustrate the fitting and error analysis techniques of the previous section we use them within a simple nuclear mass model. The model expresses nuclear binding energy as a sum of the liquid drop (LD) and shell energies [21]. The LD energy we use closely resembles the Myers-Swiatecki LD formula [22] with symmetry terms in volume and surface energy parts and a modified Coulomb part. It has the form

$$\begin{aligned} E_{\rm LD}(N,Z) &= a_V A + a_S A^{2/3} + a_{V,{\rm sym}} I^2 A \\ &+ a_{S,{\rm sym}} I^2 A^{2/3} + a_C \frac{Z(Z-1)}{A^{1/3}} + a_P \frac{P}{A^{1/2}} \,, \end{aligned} \tag{23}$$

where $I = (N-Z)/A$ and $2P = (-1)^N + (-1)^Z$. The shell energy is modelled by polynomials of $N$ and $Z$:

$$\begin{aligned} E^i_{SE}(n,z) &= x_{i,1} + x_{i,2}n + x_{i,3}z \\ &+ x_{i,4}n^2 + x_{i,5}nz + x_{i,6}z^2 \\ &+ x_{i,7}n^3 + x_{i,8}n^2z + x_{i,9}nz^2 + x_{i,10}z^3 \\ &+ x_{i,11}n^4 + x_{i,12}n^3z + x_{i,13}n^2z^2 \\ &+ x_{i,14}nz^3 + x_{i,15}z^4 \,, \end{aligned} \tag{24}$$

where $z = Z - Z_i$ and $n = N - N_i$. The index $i$ enumerates 15 different rectangular areas on the nuclear mass chart delaminated by magic numbers, see Fig. 1. In each such an area, $N$ and $Z$ values are between given magic numbers $N_i$ and $Z_i$. We restrict parameters of polynomials (24) in such a way that the shell effects be continuous across magic proton and neutron numbers, however, the derivatives thereof can be non-continuous. In this way the model can produce the binding-energy cusps at magic nucleon numbers.

The continuity requirements impose 19 conditions at semimagic nuclei, see Fig. 1. Each condition results in $p+1$ linear equations for $x_i$, where $p$ is the polynomial order. Thus for the second-, third-, or fourth-order polynomials ($p$=2, 3, or 4) we get $(p+1)\cdot 19$=57, 76, or 85 equations for 90, 150, or 225 parameters, respectively, resulting in 33, 74, or 130 independent variables of the shell energy, Eq. (24). Together with the six parameters of the liquid-drop energy, Eq. (23), the model thus contains 39, 80, or 136 independent parameters.

It should be noted that the model described above is fully linear. This means that the iteration procedure consists of just one step, because matrix $J$ is then constant and the convergence is obtained after just one iteration. In this respect the simple model considered here does not accurately resemble realistic EDF models. However, it allows us to test and showcase all the error analysis methods that can also be used in realistic nonlinear EDF calculations.

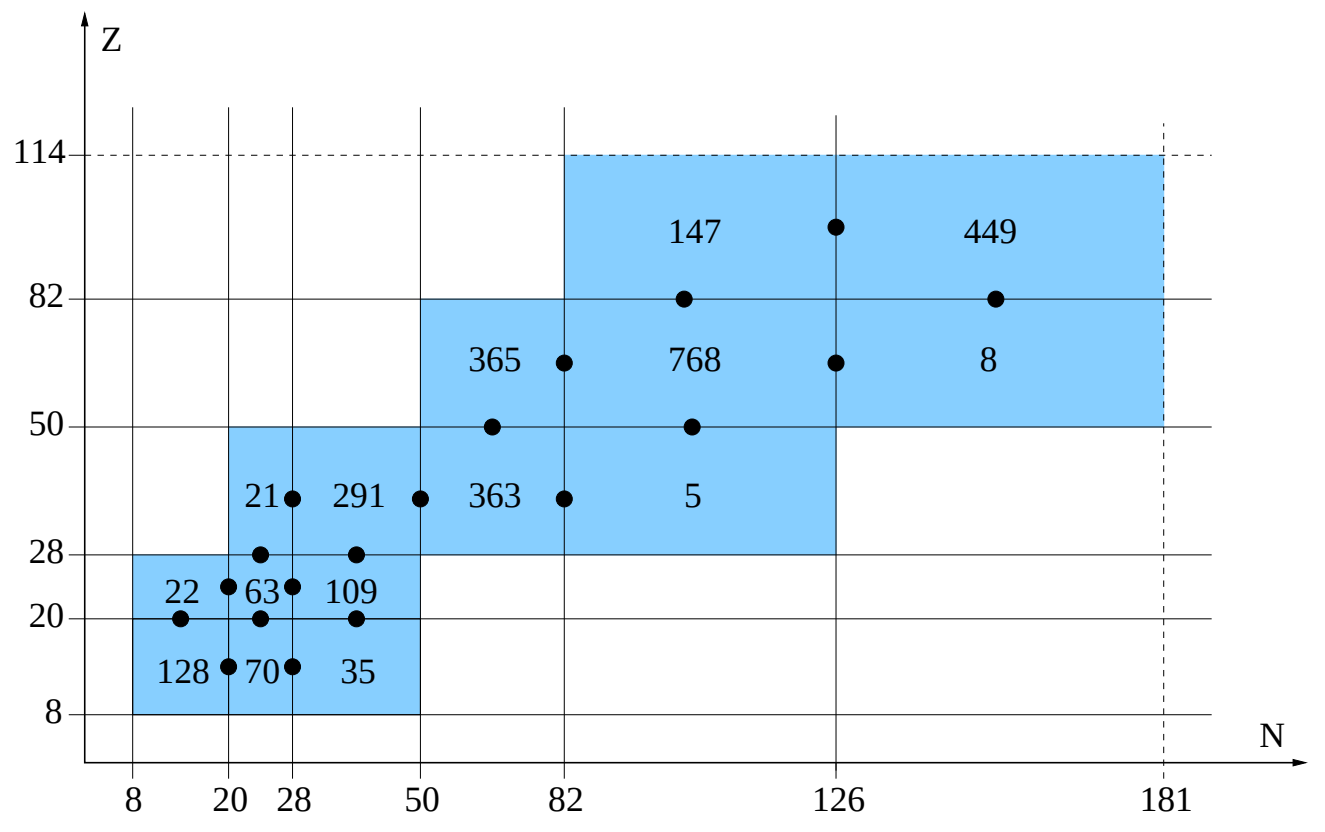

FIG. 1: (Color online) Areas of nuclear mass chart where the shell energy polynomials of Eq. (24) are defined. The black dots mark lines of semimagic nuclei for which the shell energy polynomials of adjacent rectangles are constrained to have the same values. The numbers in the rectangles show how many nuclei in the given area was used in the fit. Semimagic and magic nuclei belong always to the rectangle to the right and up.

We used the 1995 mass evaluation of Audi and Wapstra [15] as our experimental nuclear binding energies. These masses are outdated, but they serve us only for illustrative purposes. The full model with fourth-order polynomials was fitted to $m = 2844$ experimental and extrapolated binding energies of nuclei with $A \geq 16$, and the resulting set of parameters was used to create metadata masses that approximate the experimental masses with the rms deviation of 1.1 MeV. In this way, we have constructed the dataset of masses, which is exactly described by the $n = 136$ parameters of the full model. Values of the LD parameters used to define the metadata are listed in Table I.

| Parameter | Defining value | Fitted value | Error estimate |
|---|---|---|---|
| $a_V$ | 14.9455 | 14.9455 | 0.0008 |
| $a_S$ | −14.9326 | −14.9325 | 0.0024 |
| $a_{V,\mathrm{sym}}$ | −22.3303 | −22.3293 | 0.0053 |
| $a_{S,\mathrm{sym}}$ | 7.5995 | 7.5965 | 0.0068 |
| $a_C$ | −0.65709 | −0.65708 | 0.00005 |
| $a_P$ | 11.3655 | 11.3633 | 0.0187 |

TABLE I: Values and error estimates (in MeV) of the LD parameters. Values defining the metadata are compared with those obtained from fitting the exact model to metadata with the Gaussian noise of 0.1 MeV.

We do not ascribe to the model of Eqs. (23) and (24) any particular physical importance, and we are not really concerned with the question of how well it describes the experimental data. The model only serves us for the purpose of creating the metadata, and only these metadata are the subject of the consecutive analysis.

To the metadata given by the fourth-order model we add Gaussian noise of a given standard deviation $\sigma$, i.e., random numbers are added to all of the 2844 metadata masses. We stress here that we do not construct any ensemble of datasets and we do not perform any ensemble averaging. Indeed; we just have at our disposal the same number of 2844 "experimental" metadata points, for which we know exactly what are the model and noise parameters. Below, the Gaussian noise of $\sigma = 0.1$ MeV is used unless explicitly indicated.

The main thrust of our study is now at repeating the least square fits of the second-, third-, and fourth-order models described above. The fourth-order model is exact, while the second- and third-order models are inaccurate (see the discussion at the beginning of Sec II). Note that only the metadata shell effects are imprecisely described by the second- and third-order models—the LD parts of Eq. (23) have always the same form.

Our purpose is to study the fitting procedure, values of parameters, error estimates, and confidence intervals in the situations of exact and inaccurate models. In particular, we analyze dependence of the least square fits on the weights chosen for the definition of the rms deviation. To this end, we chose weights in the form

$$W_j = \frac{mA_j^\alpha}{\sum_{j=1}^m A_j^\alpha}, \tag{25}$$

where $A_j$ is the mass number of the given nuclide and $\alpha$ is a parameter. For $\alpha = 0$, one has a 'natural' choice of all weights being equal, $W_j = 1$, which is the choice most often used in nuclear mass fits.

However, it is obvious that we can equally well argue in favor of other choices. On the one hand, for $\alpha = -2$, the fit would correspond to fitting not binding energies, but binding energies par particle, $E/A$, which may seem to be a reasonable choice when discussing the LD model parameters. Naturally, this choice simply corresponds to putting a lot of more importance in masses of light than in those of heavy nuclei. On the other hand, for $\alpha > 0$, heavy nuclei are considered to be more important for the mass fits than the light ones, which can be motivated by the fact that these nuclei are closer to the infinite-matter limit. Obviously, such arguments are as good as they can get, but the bottom line is that one has here a freedom of choice that depends on personal taste and preference. Below, $\alpha$ is varied from −1 to 1, and the value of $\alpha = 0$ is used whenever not explicitly indicated.

We begin by discussing the influence of the Gaussian noise added to the metadata. In Fig. 2 we show dependence of the rms deviations of the least square fits (8) as functions of the standard deviation of the Gaussian noise $\sigma$. For the exact model, the fitting procedure reproduces perfectly well the standard deviations of the added noise. For the inaccurate models, i.e. for the second- and third-order polynomial fits, one obtains the rms deviations that are higher than the added noise.

Of course, when the added Gaussian noise goes to zero, the rms deviation of the exact model also vanishes. For inaccurate models, in this limiting case the rms deviations level out and converge to about 1.6 and 1.0 MeV for the second- and third-order models, respectively. One can say that the inaccurate models introduce their own intrinsic noises, which are not statistical in nature, but represent averaged inaccuracies of the models. One can see that at non-zero Gaussian noise, for inaccurate models the rms deviations are much smaller than the rms of the Gaussian and intrinsic noises. It looks like the intrinsic noise is gradually disappearing inside the Gaussian noise. This is in fact the limit, in which inaccurate models become quite good in describing less and less well determined experimental data.

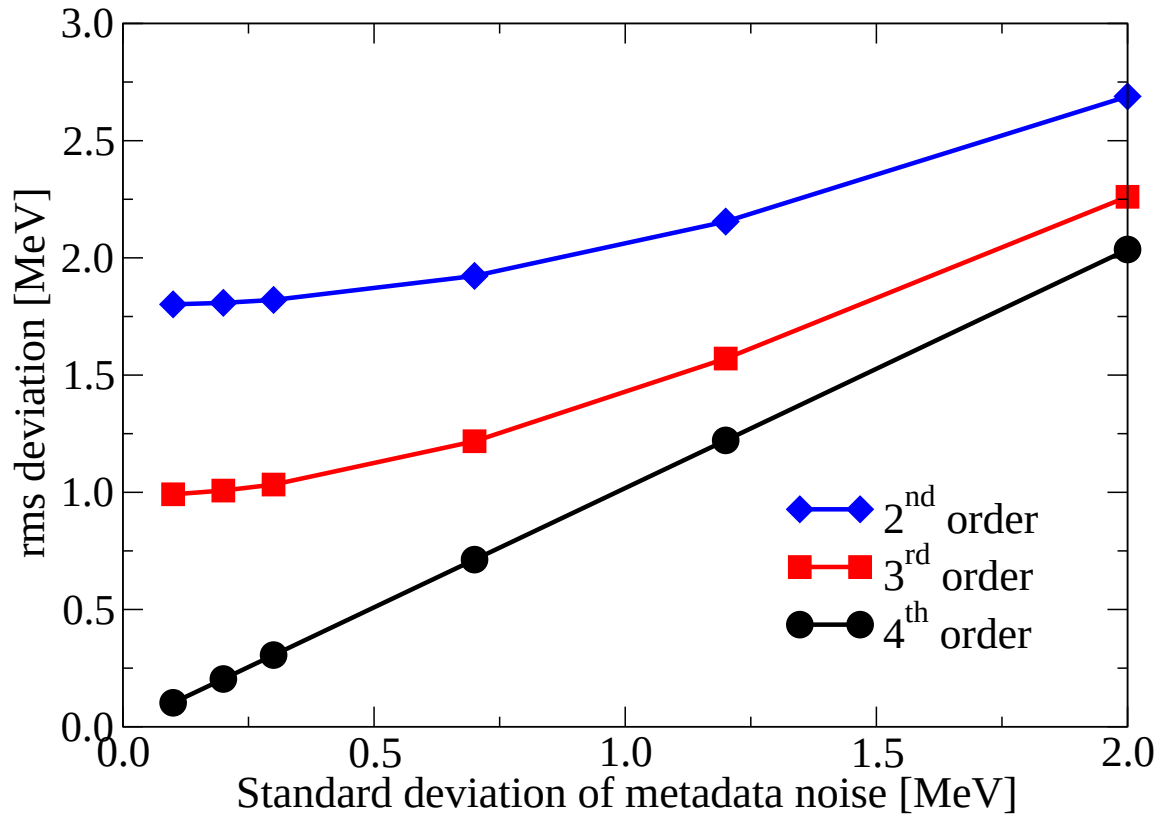

FIG. 2: (Color online) The rms deviations of the least square fits (8) as functions of the standard deviation of the Gaussian noise $\sigma$ added to the metadata.

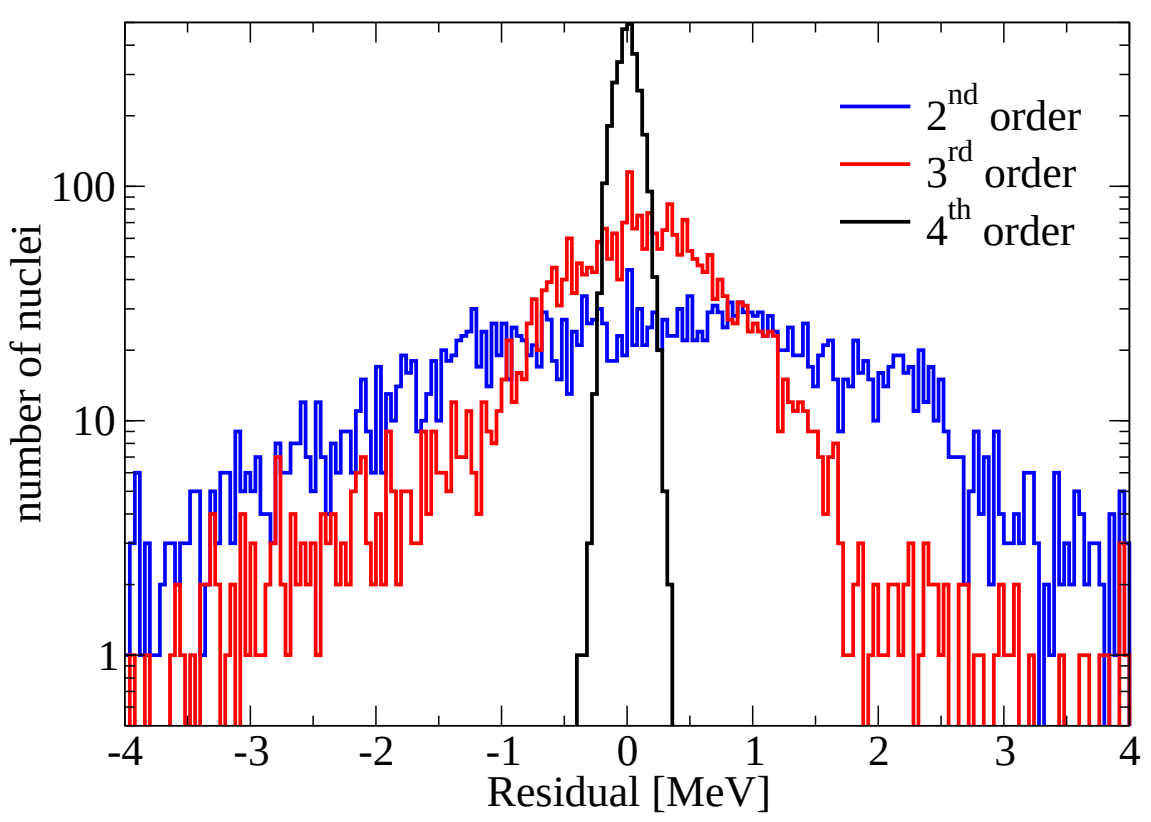

FIG. 3: (Color online) Distributions of fit residuals for three different polynomial fits to metadata. Bin widths are 0.1 MeV.

In Fig. 3 we show the distributions of fit residuals,

$$\delta e_j = e_j^{\rm exp} - f_j(\vec{x}_0), \tag{26}$$

obtained by fitting the three considered models to metadata containing the $\sigma = 0.1$ MeV Gaussian noise. As expected, for the fourth-order (exact) model, the distribution is perfectly Gaussian with the same width of 0.1 MeV. For the second- and third-order inaccurate models, the distributions are not only wider, with the widths of 1.6 and 1.0 MeV given above, but also do not have exactly Gaussian shapes. This again illustrates the non-statistical nature of the intrinsic noise within inaccurate models.

Next, we illustrate the problem of eliminating poorly determined model parameters, as explained in Eq. (14). Figure 4 shows the singular values obtained by fitting to metadata the second-, third-, and fourth-order models. When the third- and fourth-order polynomials are used in the fit, and in Eq. (14) the maximum numbers of parameters is kept, a number of parameters become ill defined. This is because some singular values of matrix $J$ become extremely small. As a result, 3 and 14 smallest singular values of the fit matrix $J$ must be eliminated when the third- and fourth-order polynomials, respectively, are used in the fits to metadata. This elimination is a direct result of some redundancy in the model parameters, which is obviously the case in those rectangles of Fig. 1 where the numbers of experimental data are small.

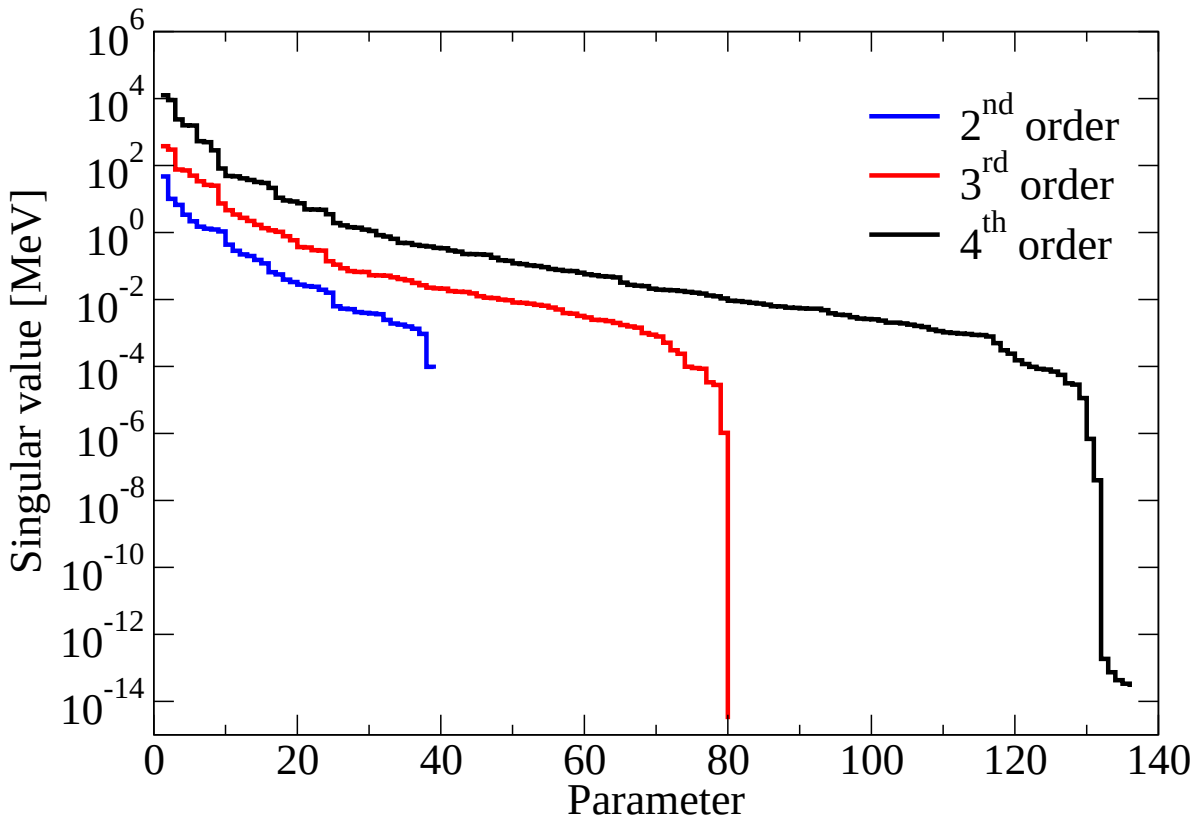

FIG. 4: (Color online) Singular values of the fit matrix $J$ of Eq. (10) when three different polynomial orders are used in the least square fit.

As can be seen from Fig. 5, even more unimportant parameters could be eliminated from the fits without losing significant amount of fit quality. If the second- or third-order polynomials are used to represent the shell effects, only about 60% of the independent uncorrelated model parameters (out of 39 or 80, respectively) are relevant and the remaining 40% do not contribute significantly to the fit, and can be safely removed. For the fourth-order (exact) model this is not the case, and many more pa-rameters (about 85% of 136) are required to go down to the value of the rms deviation equal to 0.1 MeV, corresponding to the Gaussian noise in the metadata.

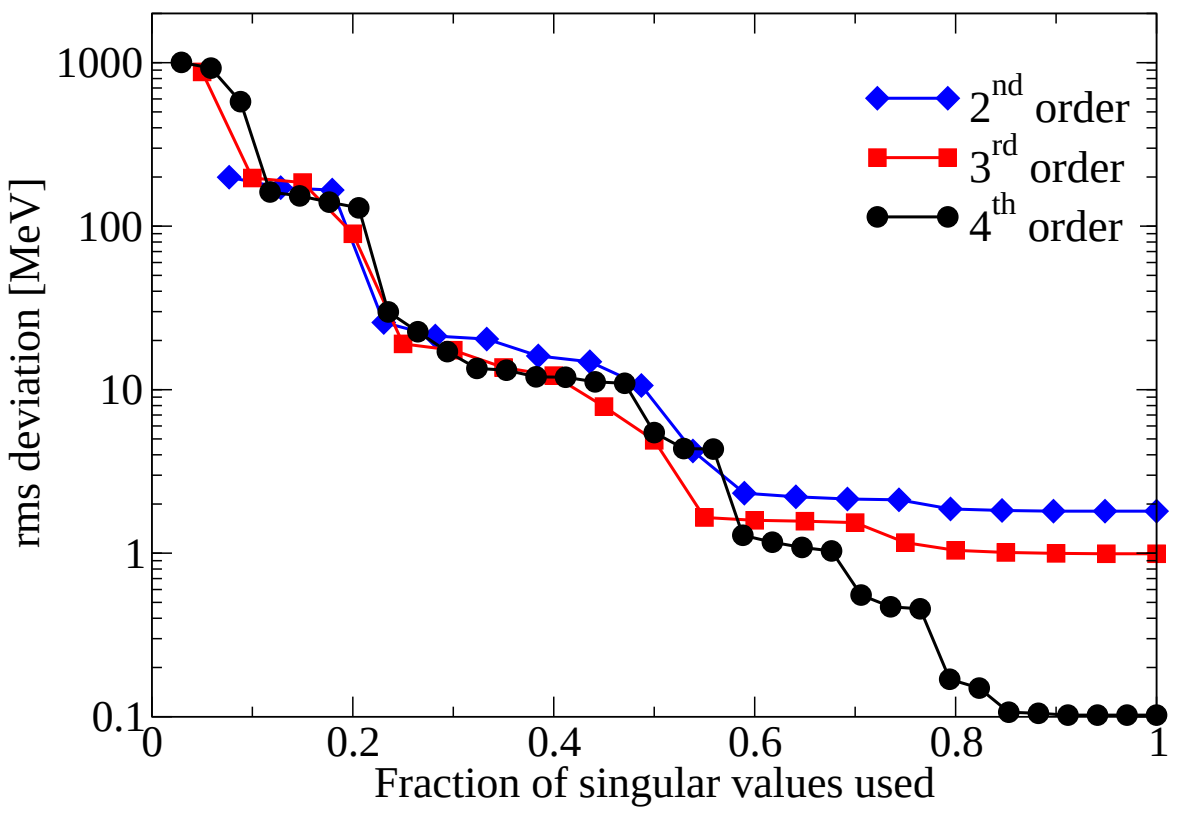

FIG. 5: (Color online) The rms deviations (8) of the least square fits to metadata, calculated for the models of Eqs. (23) and (24), as functions of number of singular values kept for matrix $J$, Eq. (14).

Figures 6 and 7 present results of fits performed for different choices of weights $W_j$, defined in Eq. (25). We first observe that fits of the fourth-order (exact) model give results that are entirely independent of weights. For $\alpha = 0$, values of fitted parameters and their error estimates are given in Table I. Small differences between the fitted values and values defining the metadata, and small values of errors, illustrate the quite small impact of the 0.1 MeV Gaussian noise included in the metadata.

Situation is drastically different for fits of the inaccurate models. Here, values of the fitted parameters, shown in Fig. 6, are not only quite different form the exact ones, but also rather strongly depend on the choice of weights. It is clear that weights strongly affect the balance between the volume and surface parameters. For weights giving greater importance to heavy nuclei ($\alpha > 0$), all absolute values of volume and surface parameters decrease. The effect is particularly large for the surface symmetry parameter $a_{S,\mathrm{sym}}$, which for the second-order model decreases from about 9 MeV at $\alpha = -1$ nearly to zero at $\alpha = 1$.

Variations of parameters, seen in Fig. 6, are much larger than their error estimates shown in Fig. 7. It means that the standard way of estimating errors, Eq. (19), may give significantly overoptimistic results. We stress here once again that the obtained variations in the LD parameters are induced by imperfect descriptions of shell effects only. One can say that such imperfections do contain smooth particle-number dependencies, which are then captured by the fitting procedure and get transferred to values of the LD parameters.

One can, in principle, argue that macroscopic (LD) and microscopic (shell) effects should not be mixed, but rather should be fitted separately to avoid cross-talk effects described above. This is certainly possible in macroscopic-microscopic models [6] that use separate expressions and/or methods to describe these two features of the mass surface. However, such separation induces ambiguities on its own, see e.g. Ref. [23], and, moreover, it cannot be realized in self-consistent methods, which describe the LD and shell effects by the same set of parameters.

In Figs. 8 and 9 we show confidence intervals and residuals, Eqs. (22) and (26), respectively, of the binding energies predicted in lead isotopes. For nuclides used in the fit (the range denoted by dotted vertical lines), confidence intervals and residuals obtained for the fourth-order (exact) model nicely reproduce the 0.1 MeV Gaussian noise included in the metadata.

Situation is again very different for the inaccurate models, which correspond to fitting the second- or third-order polynomials. In lead isotopes, residuals of the third-order model are still quite small, well below the rms deviation of 1.0 MeV, which is the value characterizing this fit. It simply means that for these observables, the model performs quite nicely. However, the confidence intervals tell us that the quality of the model even in lead nuclei is not that great as suggested by small residuals. For the second-order model, residuals become quite high but the confidence intervals indicate that the quality of the model does not, in fact, deteriorate. Confidence intervals and residuals give us diverging evaluations of quality of models, because the former represent global characteristics, which depend only on the standard deviations of parameters, while the latter illustrate only local properties of the models.

An interesting property of the confidence intervals is the fact that, for nuclei outside the fit, the confidence intervals quickly increase, independently of the complexity of the model. This result is in accordance with results obtained within realistic nuclear mass models, whose predictions (for nuclei outside the fit) deviate greatly from each other. On the one hand, such an increase of the confidence intervals is a reflection of poor predictivity of models when they are extrapolated to exotic nuclei. On the other hand, the confidence intervals simply quantify this uncertainty of extrapolation and constitute precise measures of the natural fact that such extrapolations must be uncertain. This is so because the model parameters are rather loosely defined by the metadata, and therefore, important information is missing from the models.

The discontinuity of confidence intervals at $N = 126$ is an artifact of the model, which uses different parameters in rectangles delimited by magic numbers, see Fig. 1. Note that the continuity of binding energies, but the confidence intervals need not to be continuous.

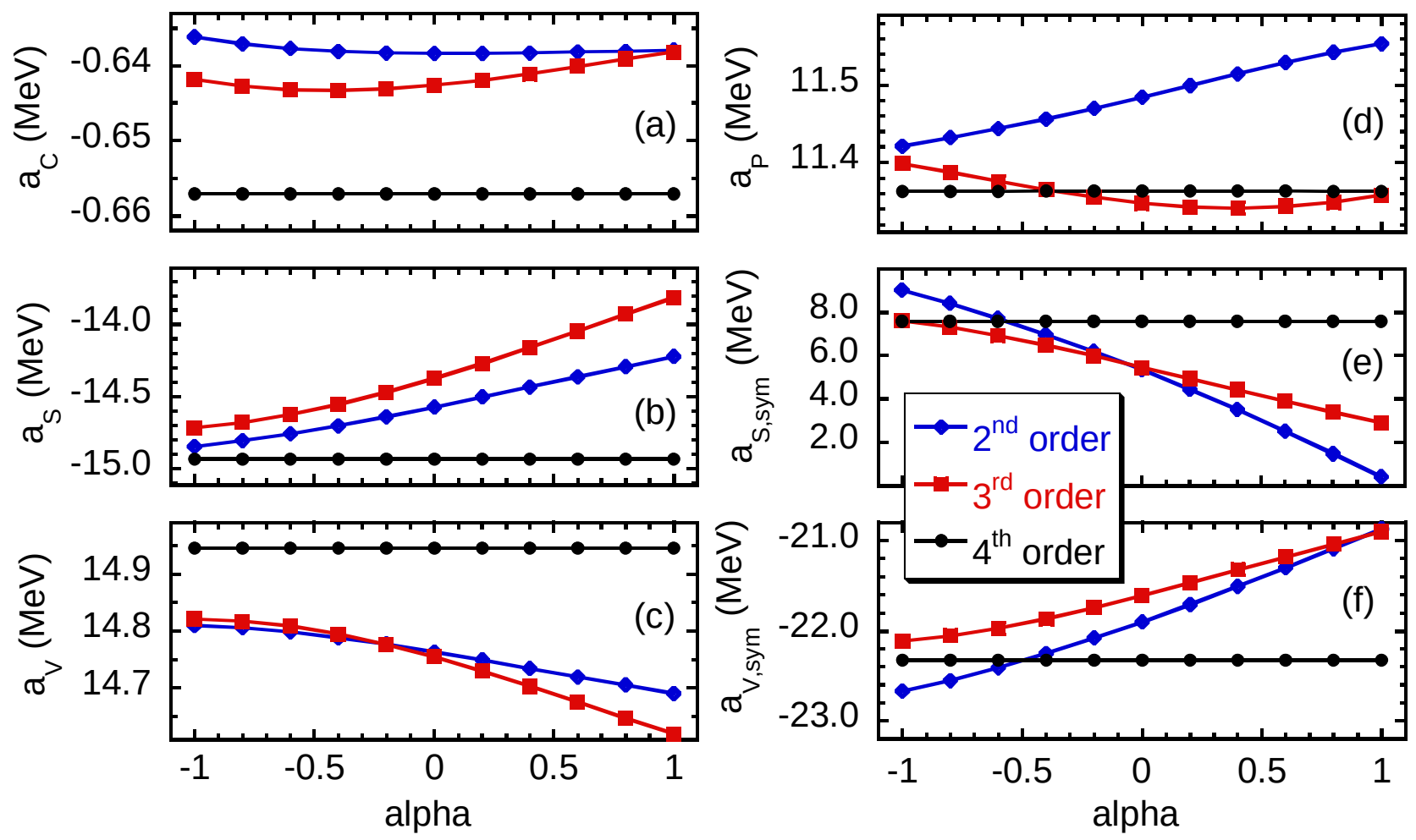

FIG. 6: (Color online) Values of the LD parameters obtained from fits with weight factors of Eq. (25), as functions of parameter $\alpha$.

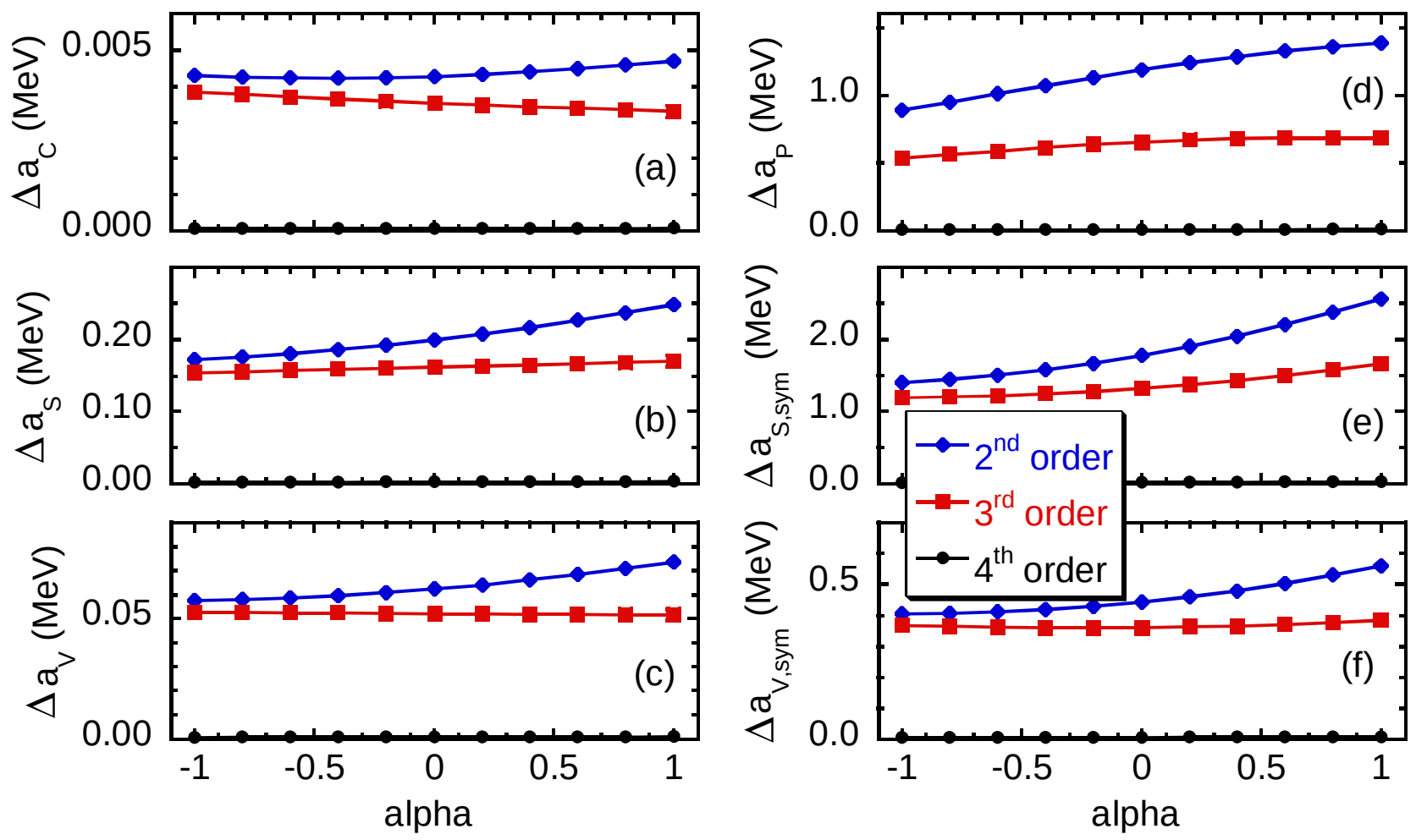

FIG. 7: (Color online) Same as in Fig. 6 but for the error estimates, Eq. (19), of the LD parameters.

## IV. CONCLUSIONS

In the present study, we have pointed out to the necessity of estimating errors along with estimating values of parameters that define nuclear mass models. Such errors allow not only for quantifying quality of models in terms of confidence intervals instead of fit residuals, but also for putting theoretical error bars on mass predictions.

A crucial element in the error analysis is the fact that the nuclear mass models belong to the class of inaccurate models, which describe data with accuracy that is much lower than that of the data themselves. For such models, standard least-square methods to estimate errors and values of parameters are not based on statistical assumptions, but rather pertain to analyzing sensitivity of the model parameters to data. Consequently, results may, and do depend on weights that are used when defining the rms deviations between the model results and data.

The discussion of error analysis was illustrated by using a simple mass model that includes a global liquid-drop part and a locally fluctuating shell-effect part, with a number of model parameters. A set of metadata masses was generated by fitting the most complex variant of the model with the fourth-order shell-effect polynomials to experimental nuclear binding energies. The metadata were then used as an "experimental" input for performing fits that used less sophisticated second- and third-order polynomials. In this way, we had at our disposal the exact model of the metadata and two inaccurate models that mimicked realistic situation in mass fits.

Within such a scheme, we were able to illustrate many properties of nuclear mass fits. In particular, we showed explicitly the relations between the statistical noise in the metadata and error estimates. We also presented meth-ods to differentiate between important and unimportant model parameters, which are based on the singular value decomposition of the regression matrix. By performing mass fits with mass-number dependent weights, we showed that values of the model parameters may involve much larger uncertainties than those given by standard error estimates. Finally, we have exemplified the role of confidence intervals and fit residuals in evaluating the quality of exact and inaccurate models.

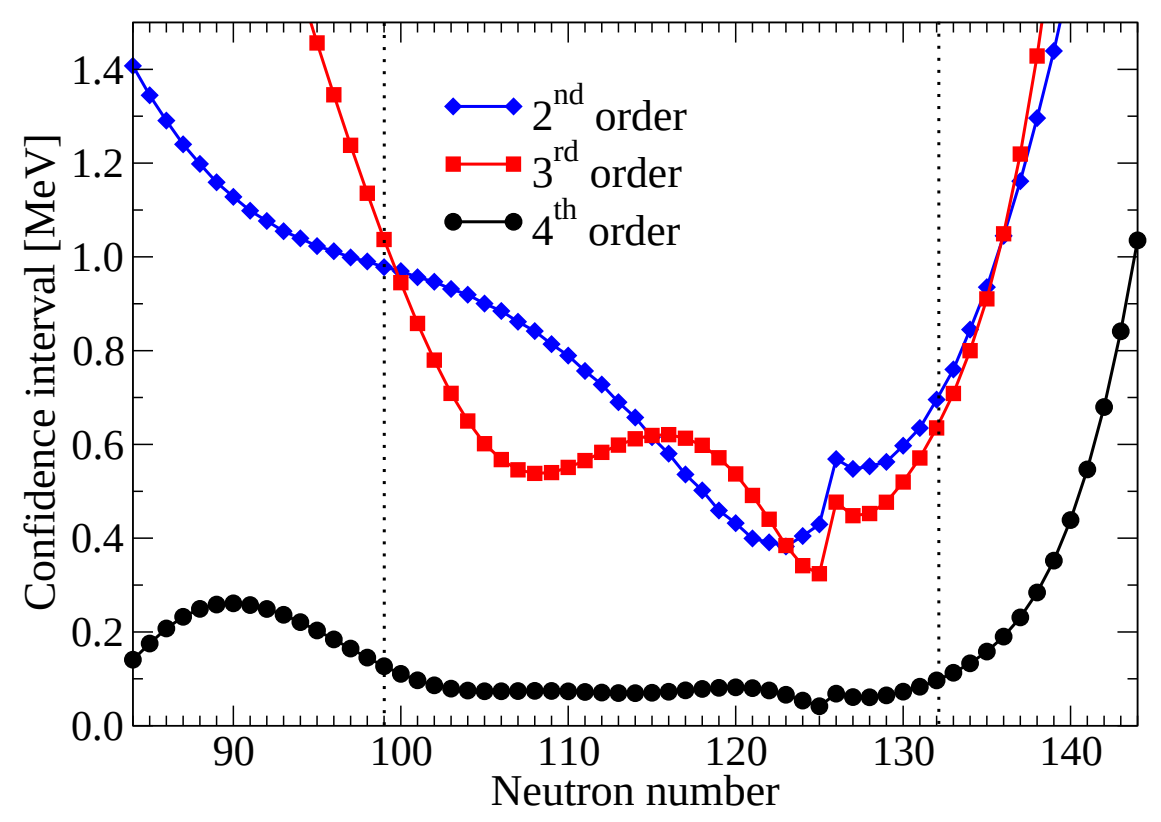

FIG. 8: (Color online) Confidence intervals (99% confidence level) of binding energies of the model defined in Eqs. (23) and (24), calculated in lead isotopes using Eq. (22).

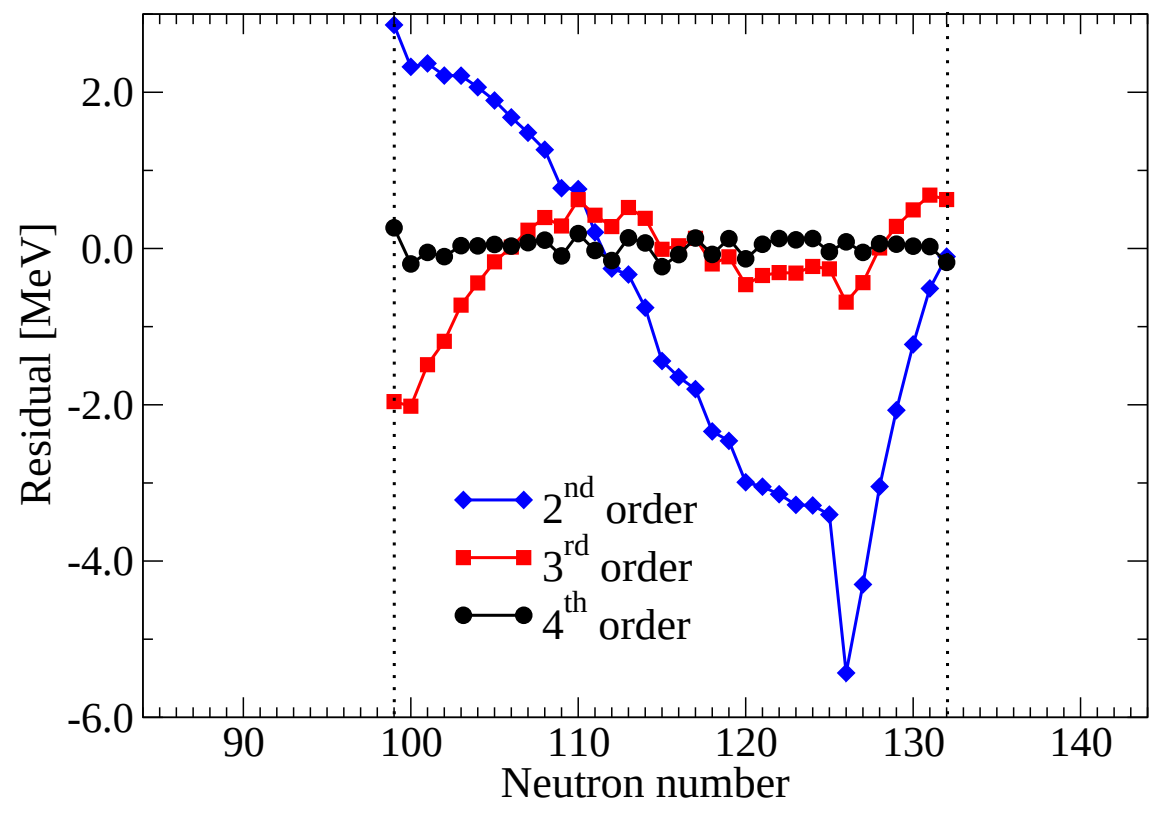

FIG. 9: (Color online) Same as in Fig. 8 but for the binding-energy residuals, Eq. (26)

Fruitful discussions with Andrzej Majhofer are gratefully acknowledged. This work was supported in part by the Academy of Finland and University of Jyväskylä within the FIDIPRO programme and by the Polish Ministry of Science under Contract No. N N202 328234.

[1] D. Lunney, J. M. Pearson, and C. Thibault Rev. Mod. Phys. **75**, 1021 (2003).
[2] S. Aoki *et al.*, Phys. Rev. **D67**, 034503 (2003).
[3] S.C. Pieper, Nucl. Phys. A **751**, 516 (2005).
[4] G.F. Bertsch, B. Sabbey, and M. Uusnäkki, Phys. Rev. C **71**, 054311 (2005).
[5] P. Klüpfel, P.-G. Reinhard, J. A. Maruhn, arXiv:0804.3402.
[6] P. Möller, J.R. Nix, W.D. Myers, and W.J. Swiatecki, Atom. Data and Nucl. Data Tables **59**, 185 (1995).
[7] S. Goriely, M. Samyn, and J. M. Pearson, Phys. Rev. C **75**, 064312 (2007).
[8] M. Bender, P.-H. Heenen, and P.-G. Reinhard, Rev. Mod. Phys. **75**, 121 (2003).
[9] J. Barea, A. Frank, J. G. Hirsch, and P. Van Isacker, Phys. Rev. Lett. **94**, 102501 (2005).
[10] M. Kortelainen, J. Dobaczewski, K. Mizuyama, and J. Toivanen, Phys. Rev. C **77**, 064307 (2008).
[11] B. Cochet, K. Bennaceur, P. Bonche, T. Duguet, and J. Meyer, Nucl. Phys. A **731**, 34 (2004).
[12] M. Kortelainen *et al.*, unpublished.
[13] G. Carlsson *et al.*, unpublished.
[14] S. Weisberg, *Applied linear regression* (Wiley-Interscience, New York, 2005).
[15] G. Audi and A.H. Wapstra, Nucl. Phys. **A595**, 409 (1995); Nucl. Phys. **A565**, 1 (1993).
[16] T. Eronen *et al.*, Phys. Rev. Lett. **97**, 232501 (2006); **100**, 132502 (2008).
[17] W.H. Press, B.P. Flannery, S.A. Teukolsky, and W.T. Vetterling, "Singular Value Decomposition", §2.6 in *Nu-*

*merical Recipes in FORTRAN: The Art of Scientific Computing* (Cambridge University Press, Cambridge, 1992).

[18] R. L. Plackett, *Principles of regression analysis*, Oxford University Press, (1960).

[19] R. F. Gunst and R. L. Mason, *Regression analysis and its application, a data oriented approach*, Marcel Dekker inc., New York (1980).

[20] W. S. Gosset (Student), Biometrika **6**, 1 (1908).

[21] P. Ring and P. Schuck, *The Nuclear Many-Body Problem* (Springer-Verlag, Berlin, 1980).

[22] W.D. Myers and W.J. Swiatecki, Ark. Fys. **36**, 343 (1967).

[23] W. Nazarewicz, T.R. Werner, and J. Dobaczewski, Phys. Rev. **C50**, 2860 (1994).